\documentclass[12pt,twoside]{article}\usepackage{amsmath} 
\input BoxedEPS
\HideDisplacementBoxes
\usepackage{subfigure}
\usepackage{latexsym}
\usepackage{amsfonts}
\SetRokickiEPSFSpecial
\usepackage{graphics}      
\usepackage{graphicx}      

\numberwithin{equation}{section}

\newcommand{\vel}{\textbf{v}}
\newcommand{\emt}{T^\mu_\nu}
\newcommand{\svel}[1]{\frac{\partial_i #1}{\partial_0 #1}}
\newcommand{\grad}{\nabla}
\newcommand{\win}{W_{\Lambda}}
\newcommand{\detmet}{\sqrt{-g}}
\newcommand{\hub}{\mathcal{H}}
\newcommand{\nlsc}{k_{\mathrm{NL}}^{-1}}
\newcommand{\sm}[1]{[#1]_{\Lambda}}

\newcommand{\pert}{\delta\rho}

\begin{document}
\title{The Effective Fluid Approach to Cosmological Nonlinearities: Applications to Preheating}\author{Hyeyoun Chung\\\small \it{Jefferson Physical Laboratory, Harvard University,}\\\small\it{17 Oxford St., Cambridge, MA 02138, USA}\\\small hyeyoun@physics.harvard.edu\\[-0.15in]} \date{\small\today} 


\maketitle
\begin{abstract}\noindent 
In \cite{Baumann}, Baumann et al. presented a new formalism for studying cosmological systems where the characteristic scale of non-linearities, $\nlsc$, is much smaller than the Hubble scale $\hub^{-1}$. By integrating out the short-wavelength modes, it is possible to obtain an effective theory of long-wavelength perturbations that is described by an imperfect fluid evolving in an FRW background. As the long-wavelength perturbations remain small even when the short-scale dynamics are non-linear, the tools of linear perturbation theory may be applied. The work in \cite{Baumann} deals only with matter in the form of a pressureless perfect fluid with zero anisotropic stress, and also assumes that the short-scale gravitational dynamics are Newtonian. In this work we extend this formalism to the case of a perfect fluid with pressure, and in particular to the case of preheating after inflation, where the matter content of the universe can be modeled by two coupled scalar fields. We discard the assumption that the short-scale gravitational dynamics are Newtonian. We find that our results differ from Baumann et al.'s even when the pressure is set to zero, which suggests that relaxing their assumptions creates appreciable changes in the long-wavelength effective theory. We derive equations of motion for the total density perturbation and matter velocities during preheating, as well as linearized Einstein equations for the long-wavelength metric perturbations. We also present the equations governing the effective long-wavelength scalar field dynamics.
\end{abstract}

\section{Introduction}

Recently, Baumann et al. have proposed a formalism for analytically studying long-wavelength dynamics even when the density contrast $\delta$ of a universe grows large\cite{Baumann}. Their method applies when the characteristic scale of non-linearities, $\nlsc$, is much smaller than the Hubble scale $\hub^{-1}$. This hierarchy allows us to integrate out the short-wavelength modes by smoothing all perturbations over a scale $\Lambda^{-1}$ that lies between the non-linear scale and the scale of the long-wavelength perturbations. This procedure gives an effective theory of long-wavelength perturbations that is modeled to lowest order by an imperfect fluid with \textit{effective energy-momentum pseudotensor} $\tau_{\mu\nu}$, evolving in an FRW universe. The properties of the effective fluid are determined by the interactions of the short-wavelength modes. 

The equations of motion in the effective theory can be expressed as a derivative expansion of the long-wavelength variables, with higher order terms being suppressed by powers of $(k/\Lambda)^2$ or $(k/k_{NL})^2$. Since the long-wavelength perturbations during preheating remain small even when the underlying dynamics have become non-linear, we can then apply perturbation theory techniques. The coefficients of the effective theory can either be obtained by matching to the results of numerical simulations, or left as free parameters to be matched to measurements from experiment.

Baumann et al. considered a universe filled with cold dark matter, modeled as a pressureless perfect fluid. In this paper we consider a universe where the matter content is a perfect fluid with pressure: in particular, we consider matter in the form of two coupled scalar fields. We also make fewer assumptions than Baumann et al. For example, we discard their assumption that the short-scale gravitational dynamics are Newtonian. We obtain somewhat different results from those given in \cite{Baumann}, even when we set the pressure to zero in our equations, suggesting that relaxing the underlying assumptions on the short-scale gravitational dynamics significantly affects the long-wavelength perturbations. We hope that our approach will thus be applicable in more general scenarios than a $\Lambda$CDM universe.

One particular example where our work applies is the period of preheating following inflation. Inflation ends when the slow-roll conditions are violated, and the inflaton $\phi$ begins to oscillate around its ground state. As the inflaton field is coupled to the standard model (SM) matter fields, these oscillations cause the energy stored in $\phi$ to be transferred to the SM fields. This process is known as \textit{reheating}\cite{Bassett}.

Reheating was originally analyzed using perturbative quantum field theory, until it was realized that the coherent nature of the inflaton field at the end of inflation renders this picture inaccurate. The excitation of SM fields during reheating was then reformulated as a semi-classical problem, in which the quantum mechanical production of SM matter particles takes place in the classical background of the inflaton field. The most common toy model used to describe this situation couples a scalar matter field $\chi$ to the inflaton field $\phi$. In this model of reheating the $\chi$ field is excited via parametric resonance, and the inflaton decays rapidly as a result, so that the dynamics quickly become non-linear. In this regime the standard tools of linearized cosmological perturbation theory no longer apply, so most studies of reheating have relied on numerical simulations\cite{DEFROST}. In particular, the occupation numbers of $\chi$ and $\phi$ are found to be large in the non-linear domain, so that they may be treated as classical fields and studied using lattice simulations. This period of rapid energy transfer from $\phi$ to $\chi$ is known as \textit{preheating}. It results in a highly non-thermal distribution of matter fields that then thermalizes to give the initial conditions for a hot big bang.

Although computer simulations are useful for providing visual representations of preheating dynamics, it would be desirable to have an analytical description of the non-linear dynamics. In addition to allowing us to make explicit calculations, it could also offer a greater understanding of the scalar field interactions. As preheating is a highly inhomogeneous process, the density contrast $\delta \equiv \rho/\bar{\rho} - 1$ for both $\phi$ and $\chi$ soon exceeds 1, so that conventional cosmological perturbation theory (in which we expand in $\delta$, matter velocity $\vel$, and metric perturbations $\Phi$) cannot be applied. However, numerical studies have shown that there is a large hierarchy between the non-linear scale and the Hubble scale during preheating, with $k_{\mathrm{NL}}^{-1} \sim 0.01\mathcal{H}^{-1}$\cite{DEFROST}. Moreover, the metric and velocity perturbations remain small, even while the inhomogeneous parts of $\phi$ and $\chi$ grow large\cite{Frag}. Thus it seems that the period of preheating provides a suitable test case for setting up an effective theory of long-wavelength perturbations.

This paper is organized as follows. In Section \ref{sec-Exp} we describe the rules for expanding the equations of motion in our formalism. In Section \ref{sec-Equations} we give the basic equations that govern the scalar field and metric dynamics during preheating. In Section \ref{sec-Long} we derive the effective theory of the long-wavelength modes, and in Section \ref{sec-Results} we give the evolution equations for the long-wavelength theory. We conclude in Section \ref{sec-Discussion}.

\section{Cosmological Perturbation Theory: A Velocity Expansion}\label{sec-Exp}

In cosmological perturbation theory, we usually expand to linear order in the density perturbation $\delta\rho$, matter velocity $\textbf{v}$, and metric perturbations $\Phi$, so that linear perturbation theory is no longer valid when the density contrast $\delta > 1$. In our work, we follow the alternative approach outlined in \cite{Baumann}, where we expand up to order $v^2$ in velocity and metric perturbations, and we do \textit{not} expand in $\pert$.

In perturbation theory, we find that $\vel$ is related to the Newtonian potential $\Phi$ and the density contrast $\delta$ by
\begin{equation}
v^2 \sim \Phi\delta
\end{equation}
In the non-linear regime, when $\delta \sim 1$, we therefore find that $v^2\sim\Phi$.

On small scales, gradients of the gravitational potential can change the power-counting of standard perturbation theory, as the short-wavelength modes have large momenta $k$. The net result is that, at the non-linear scale, each gradient of $\Phi$ reduces the order in $v$ by one. Therefore, when we carry out an expansion to order $v^2$, we expand to linear order in $\Phi$ and second order in $\grad\Phi$.

Simulations of the scalar fields during preheating have shown that the metric fluctuations and the gradient energies of the fields remain small, even when the fields themselves (and their kinetic energies) grow large\cite{Frag}. Therefore, it is valid to apply this expansion when studying preheating dynamics.

\section{Preheating: The Basic Equations}\label{sec-Equations}

We will consider one of the simplest models of preheating, in which a massive inflaton field $\phi$ with inflaton potential $V_1=\frac{1}{2}m^2\phi^2$ couples to a massless scalar field $\chi$ through a potential $V_2 = \frac{1}{2}g^2\phi^2\chi^2$. (Our approach can easily be extended to other models with different interaction potentials.) The full potential is therefore
\begin{equation}
V(\phi,\chi)=V_1 + V_2 = \frac{1}{2}m^2\phi^2 +\frac{1}{2}g^2\phi^2\chi^2
\end{equation}
At the end of inflation $\phi$ is a homogeneous field that approaches
\begin{equation}
\phi \rightarrow \frac{M_{pl}}{\sqrt{3\pi}mt}\sin mt,
\end{equation}
where $t$ is the cosmic time. The spacetime is described by a perturbed FRW metric
\begin{equation}
\mathrm{d}s^2 = a^2(\eta)\left (-e^{2\Psi}\mathrm{d}\eta^2 + e^{-2\Phi}\mathrm{d}\textbf{x}^2 \right ),
\end{equation}
where $\eta$ is the conformal time, we have ignored vector and tensor perturbations, and we use the Poisson gauge. In the case of zero anisotropic stress, we find that $\Psi = \Phi$ to first order.

The basic equations that describe the dynamics during preheating are the Klein-Gordon (KG) equations for $\phi$ and $\chi$, and the Einstein equations. The KG equations are:
\begin{align}
\Box \phi - V_{,\phi} &= 0\\
\Box \chi - V_{,\chi} &= 0
\end{align}
where $V_{,\phi}\equiv\frac{\partial V}{\partial\phi}$. The energy-momentum tensor of the scalar fields is given by
\begin{align}
\emt &= \partial_\mu\phi\partial^\nu\phi +\partial_\mu\chi\partial^\nu\chi - \delta^\mu_\nu\left (\frac{1}{2}\partial_\alpha\phi\partial^\alpha\phi+\frac{1}{2}\partial_\alpha\chi\partial^\alpha\chi+V \right )\label{eq-EMT}
\end{align}
Furthermore, $\emt$ obeys the conservation law
\begin{equation}
\grad_\mu\emt = 0.\label{eq-EMTCons}
\end{equation}

Finally, the Einstein equations are given by
\begin{equation}
G^\mu_\nu = 8\pi G \emt.
\end{equation}

In the rest of this section we expand the KG equations and the continuity equations to order $v^2$ in the metric perturbations and the matter velocities, but we do not expand in $\rho$ or $P$. We find that the Einstein equations have a very similar form to those given in \cite{Baumann}, at least to leading order in long-wavelength perturbations. We therefore give these equations in Appendix \ref{app-EE}. In order to apply the expansion to the system of coupled scalar fields described above, we first consider the ways in which this system may be interpreted as a sum of perfect fluids.

\subsection{The Scalar Field as a Perfect Fluid}

It is possible to interpret a system of interacting scalar fields as a system of interacting fluids \cite{Corresp, Malik, Madsen, Bruni}. We can therefore expand the equations of motion in terms of $\rho$ and $\vel$, or in terms of $\phi$ and $\chi$. This will give us two equivalent descriptions of the matter fields during preheating, and we can go back and forth between the two using the correspondence outlined below. It will be useful to consider both descriptions, as they offer different physical insights and different calculational advantages. The fluid description allows us to consider the energy density of the entire system using only one variable, $\rho$, and gives a simpler formulation of the equations of motion. The scalar field description allows us to consider the individual field perturbations directly.

The energy momentum tensor of a perfect fluid with energy density $\rho$, pressure $P$, and zero anisotropic stress is given by
\begin{equation}
\emt = (\rho + P)u^\mu u_\nu + \delta^\mu_\nu P,\label{eq-EMTFluid}
\end{equation}
where $u^\mu$ is the instantaneous 4-velocity of the fluid. The components of the 4-velocity are related to the matter velocity $\vel$ by the equations
\begin{align}
u^0 &=a^{-1}e^{-\Psi}\gamma(v), &u^i=a^{-1}e^\Phi v^i\\
u_0 &=-a e^{\Psi}\gamma(v), &u_i=ae^{-\Phi}v^i,
\end{align}
where $\gamma(v) := (1-v^2)^{-1/2}$. Comparing Eq.(\ref{eq-EMTFluid}) with the form of $\emt$ in Eq.(\ref{eq-EMT}), we see that the interacting scalar fields $\phi$ and $\chi$ can be treated as the sum of two ``kinetic fluids'' with energy density and pressure\cite{Malik}
\begin{align}
\rho_{\phi} &:= -\frac{1}{2}\partial_\alpha\phi\partial^\alpha\phi &\rho_{\chi} &:= -\frac{1}{2}\partial_\alpha\chi\partial^\alpha\chi\\
P_{\phi} &:= -\frac{1}{2}\partial_\alpha\phi\partial^\alpha\phi &P_{\chi} &:= -\frac{1}{2}\partial_\alpha\chi\partial^\alpha\chi
\end{align}
and a single ``potential fluid'' with energy density and pressure
\begin{align}
\rho_V &:= V\\
P_V &:= -V.
\end{align}
The instantaneous 4-velocity of each kinetic fluid is is given by
\begin{align}
u_{\phi}^\mu &:= \frac{-\partial^\mu\phi}{\sqrt{-\partial_\alpha\phi\partial^\alpha\phi}} &u_{\chi}^\mu &:= \frac{-\partial^\mu\chi}{\sqrt{-\partial_\alpha\chi\partial^\alpha\chi}}
\end{align}
and to order $v^2$ the matter velocity $\vel$ is given by
\begin{align}
v_{\phi}^i &=\frac{\partial^i\phi}{\partial^0\phi} = -\frac{\partial_i\phi}{\partial_0\phi}, &v_{\chi}^i &=\frac{\partial^i\chi}{\partial^0\chi} = -\frac{\partial_i\chi}{\partial_0\chi}
\end{align}
The 4-velocity is not defined for the potential fluid.

The total energy-momentum tensor is given by the sum of the individual energy-momentum tensors of these three fluids. The total veocity perturbation is given by
\begin{align}
v^i &= \frac{(\rho_\phi + P_\phi)v^i_\phi+(\rho_\chi + P_\chi)v^i_\chi}{\rho+P}
\end{align}
where $\rho := \rho_\phi + \rho_\chi + \rho_V$ and $P := P_\phi + P_\chi + P_V$ are the total energy density and total pressure respectively.

The physical interpretation of the scalar field as a fluid is valid as long as $u^\mu$ is timelike. In our case, we are assuming that $\vel$ is small compared to $\delta$, which is equivalent to assuming that the gradient energy of the scalar fields is small compared to their kinetic energy. Thus it is reasonable to assume that $u^\mu$ remains timelike during preheating. The form of $\emt$ shows that the anisotropic stress is zero. Therefore, we can set the metric perturbations $\Phi = \Psi$ to first order, and will do so for the rest of the paper.

\subsection{The Energy-Momentum Conservation Equations}\label{sec-TExp}

The components of the energy-momentum tensor calculated to order $v^2$ in terms of $\rho$, $\vel$ are:
\begin{align}
T^0_0 &=-(\rho + P)\gamma^2 +P\label{eq-EMT00}\\
T^i_0 &=-v^i(\rho+P)\label{eq-EMT01}\\
T^i_j &=(\rho+P)v^iv^j+ \delta^i_j P\label{eq-EMTij}
\end{align}
In terms of $\phi$, $\chi$, the components are:
\begin{align}
T^0_0 &=-\frac{1}{2a^2}\left ( 1-2\Psi-\left(\svel{\phi}\right)^2\right ) (\partial_0\phi)^2 -\frac{1}{2a^2}\left ( 1-2\Psi-\left(\svel{\chi}\right)^2\right ) (\partial_0\chi)^2\nonumber\\
&\,\,\,\,\,\,\,\,\,- V\\
T^i_0 &=\frac{1}{a^2}\left (\partial_i\phi\partial_0\phi+\partial_i\chi\partial_0\chi\right )\\
T^i_j &=\frac{1}{a^2}\left (\partial_i\phi\partial_j\phi+\partial_i\chi\partial_j\chi\right )-\delta^i_jV\nonumber\\
&+\frac{1}{2a^2}\delta^i_j\left (\left ( 1-2\Psi-\left(\svel{\phi}\right)^2\right ) (\partial_0\phi)^2+\left ( 1-2\Psi-\left(\svel{\chi}\right)^2\right ) (\partial_0\chi)^2\right )
\end{align}
In \cite{Baumann}, the equations of motion of the long-wavelength perturbations are derived using the Euler equations in the Newtonian approximation. We will take a different approach and use the full general-relativistic conservation equation (\ref{eq-EMTCons}). We can project this equation along $u^\nu$, or along the orthogonal direction:
\begin{align}
u^\nu\grad_\mu\emt &= -\frac{1}{\detmet}\partial_\mu\left (\detmet (\rho + P)u^\mu \right ) + u^\mu\partial_\mu P = 0\label{eq-Proj1}\\
(g^{\sigma\nu} + u^\sigma u^\nu)\grad_\mu\emt &= (\rho + P)u^\mu \grad_\mu u^\sigma + \partial^\sigma P + u^\sigma u^\mu\partial_\mu P = 0\label{eq-Proj2}
\end{align}
To order $v^2$, Eq.(\ref{eq-Proj1}) is:
\begin{equation}
\dot{\rho} + \partial_i((\rho+P)v^i) - v^i\partial_iP + (3\hub -3\dot{\Phi}+2\vel\cdot\dot{\vel})(\rho+P)\label{eq-Cons1}
\end{equation}
To order $v^2$, the spatial components of Eq.(\ref{eq-Proj2}) are:
\begin{align}
(\rho + P)\left ( \dot{v_i} + v^j\partial_jv_i + \partial_i\Psi\right ) + \partial_i P + v_i\partial_0 P + v_iv^j\partial_j P = 0\label{eq-Cons2}
\end{align}

\subsection{The Einstein Equations}\label{sec-EE}

To write down the Einstein equations to order $v^2$, we decompose the Einstein tensor $G_{\mu\nu}$ into a homogeneous background $\bar{G}_{\mu\nu}$, a part linear in the perturbations ($G_{\mu\nu}^L$), and part non-linear in the perturbations ($G_{\mu\nu}^{NL}$). We can then write the non-linear Einstein equations in the form
\begin{align}
G_{\mu\nu}^L &= 8\pi G(\tau_{\mu\nu} - \bar{T}_{\mu\nu})
\end{align}
where the energy-momentum pseudotensor $\tau_{\mu\nu}$ is given by
\begin{equation}
\tau_{\mu\nu} \equiv T_{\mu\nu} - \frac{G_{\mu\nu}^{NL}}{8\pi G}\label{eq-EMTPseudo}
\end{equation}
The velocity expansion of the Einstein equations is given in Appendix \ref{app-EE1}.

\subsection{The Klein-Gordon Equations}

Expanding the Klein-Gordon equations for $\phi$to order $v^2$ gives:
\begin{align}
(1-2\Psi)\partial_0^2\phi - \partial_i^2\phi+(2\hub - 4\Psi\hub - 3\dot{\Phi} - \dot{\Psi})\partial_0\phi +\partial_i(\Phi - \Psi)\partial_i\phi+V_{,\phi} &= 0,
\end{align}
with an analogous equation holding for $\chi$.

\section{Integrating Out Short-Wavelength Modes}\label{sec-Long}

In this section we describe how to integrate out the short-wavelength modes to obtain a long-wavelength effective theory. Integrating out the short-wavelength modes amounts to averaging perturbations over a smoothing scale $\Lambda^{-1}$. Since we are interested in the theory at scales $k^{-1}$ much larger than the non-linear scale $k^{-1}_{\mathrm{NL}}$, we choose a smoothing scale $\Lambda^{-1} >> k_{\mathrm{NL}}^{-1}$.

The smoothing of perturbations corresponds to a convolution of all fields $X \equiv \{ \rho, \Phi, \rho\vel\}$ with a window function $\win$. We define the long-wavelength mode $X_l$ of a field $X$ to be
\begin{equation}
X_l \equiv [X]_{\Lambda}(\textbf{x}) = \int \mathrm{d}^3\textbf{x}'\,\, \win(|\textbf{x}-\textbf{x}'|)X(\textbf{x}').
\end{equation}
The short-wavelength mode $X_s$ of $X$ is then defined by
\begin{equation}
X \equiv X_l + X_s.
\end{equation}
We will assume $\win$ to be Gaussian for convenience. We also assume that $\win$ satisfies the following conditions:
\begin{align}
\partial_{j'}\win &= -\partial_j\win = \Lambda^2(\textbf{x} - \textbf{x}')^j\win\\
\partial_{i'}\partial_{j'}\win &= \partial_i\partial_j\win = -\Lambda^2\delta_{ij}\win + \Lambda^4(\textbf{x} - \textbf{x}')^i(\textbf{x} - \textbf{x}')^j\win
\end{align}
If we smooth general bilinear and trilinear quantities using $\win$, we obtain the following results:
\begin{align}
\sm{fg} &= f_lg_l + \sm{f_sg_s} + \frac{1}{\Lambda^2} \grad f_l\cdot \grad g_l + \dots\label{eq-bi}\\
\sm{fgh} &= f_lg_lh_l + \sm{f_sg_sh_s}+f_l\sm{g_sh_s}+g_l\sm{f_sh_s}+h_l\sm{f_sg_s}\nonumber\\
&+ \frac{1}{\Lambda^2}(f_l \grad g_l\cdot \grad h_l+g_l \grad f_l\cdot \grad h_l+h_l \grad f_l\cdot \grad g_l)\nonumber\\
&+ \frac{1}{\Lambda^2}(\grad f_l \cdot \grad\sm{g_sh_s}+\grad g_l \cdot \grad\sm{f_sh_s}+\grad h_l \cdot \grad\sm{f_sg_s})\nonumber\\
&+ \frac{1}{2\Lambda^2}(\grad^2 f_l \sm{g_sh_s}+\grad^2 g_l \sm{f_sh_s}+\grad^2 h_l \sm{f_sg_s})+ \dots\label{eq-tri}
\end{align}
The explicit calculations used to derive Eq.(\ref{eq-bi}-\ref{eq-tri}) are given in Appendix \ref{sec-App1}. The expressions are given up to higher derivative terms of order $k^2/\Lambda^2$, where $k$ is a characteristic frequency of a long-wavelength mode. The higher derivative terms are therefore suppressed, as $k<<\Lambda$. As mentioned in the Introduction, our expression for smoothed trilinear quantities differs from that given in \cite{Baumann}, due to the presence of extra terms that are absent in their formula. It is possible that the discrepancy is due to extra (unspecified) assumptions that have been imposed in their paper to allow these terms to be dropped.

\section{The Long-Wavelength Effective Theory}\label{sec-Results}

We now use the equations from Section \ref{sec-Equations} and Section \ref{sec-Long} to derive the long-wavelength effective 
theory of preheating dynamics. Smoothing the energy-momentum pseudotensor in (\ref{eq-EMTPseudo}), we find that the \textit{effective} energy-momentum pseudotensor $\sm{\tau_{\mu\nu}}$ has the form
\begin{equation}
\sm{\tau_{\mu\nu}} = \tau_{\mu\nu}^l + \tau_{\mu\nu}^s + \tau_{\mu\nu}^{\partial^2},
\end{equation}
where $\tau_{\mu\nu}^l$ depends only on the long-wavelength perturbations, $\tau_{\mu\nu}^s$ depends on the short-wavelength modes, and $\tau_{\mu\nu}^{\partial^2}$ contains higher-derivative corrections that are suppressed by powers of $k^2/\Lambda^2$. Throughout this work, we will drop all such higher-derivative corrections that result from smoothing. We also drop non-linear metric contributions to $\tau_{\mu\nu}^l$.

The pseudotensor $\sm{\tau_{\mu\nu}}$ describes an imperfect fluid. Thus the effective theory we obtain after smoothing is an imperfect fluid with density perturbation $\delta_l$ and velocity perturbation $\vel_l$, evolving in a background FRW metric with scalar metric perturbations $\Phi_l$ and $\Psi_l$. As the scale of non-linearities is much smaller than the smoothing scale, these long-wavelength perturbations remain small even when the small-scale dynamics have become non-linear. Therefore, we may apply linear perturbation theory in the effective theory even when $\delta>>1$ at small scales.

To leading order in the long-wavelength perturbations, we may write the energy-momentum pseudotensor as:
\begin{equation}
\sm{\tau_{\mu\nu}} = (\rho + P)u_\mu u_\nu + (P-\zeta\theta)g_{\mu\nu} +\Sigma_{\mu\nu}\label{eq-TAnsatz}
\end{equation}
where $\theta = \partial_iu_i$, $\zeta$ is the bulk viscosity, and $\Sigma^\mu_\nu$ is the anisotropic stress. The pseudotensor is given as a derivative expansion, with higher order terms being suppressed by $(k/k_{NL})^2$. All of the quantities in (\ref{eq-TAnsatz}) are quantities in the effective theory: for example, $\rho := \rho_{\mathrm{eff}} = \bar{\rho}_{\mathrm{eff}} + \delta\rho_{\mathrm{eff}}$. In order to avoid cluttering the notation, we will omit the ``eff'' subscript from relevant quantities, with the exception of Section \ref{sec-Back} when we discuss the renormalization of background pressure, energy density, and anisotropic stress due to short-wavelength dynamics. Although we began with a system of coupled scalar fields that had zero anisotropic stress and zero viscosity, we will find that both anisotropic stress and viscosity are induced in the long-wavelength effective theory.

We use the following ansatz for the anisotropic stress:
\begin{equation}
\Sigma_{ij} = -\eta\sigma_{ij},\hspace{3cm} \sigma_{ij} := v_{(i,j)} - \frac{1}{3}\delta_{ij}v_{k,k},
\end{equation}
where $\eta$ is the shear viscosity. To lowest order, the pressure perturbation $\delta P$ is
\begin{equation}
\delta P = c_s^2 \rho\delta,
\end{equation}
where $c_s^2$ is the speed of sound squared. We also define the equation of state parameter $w = \frac{P}{\rho}$, and the dimensionless paramter $c_{vis}^2$ that characterizes the viscosity by
\begin{equation}
c_{vis}^2:= \left (\frac{2\eta}{3}+\zeta\right)\frac{\hub}{\bar{\rho}}
\end{equation}
The anisotropic stress induced in the effective theory satisfies
\begin{equation}
-\frac{1}{\bar{\rho}}\frac{k_ik_j}{k^2}\Sigma^i_j = -c_{vis}^2\frac{\theta}{\hub}
\end{equation}
To leading order in the long-wavelength perturbations, we have
\begin{align}
\frac{1}{\bar{\rho}}\frac{k_ik_j}{k^2}\sm{\tau_{ij}}^s = c_s^2\delta - c_{vis}^2\frac{\theta}{\hub}
\end{align}

\subsection{The Effective Fluid}\label{sec-Back}

At super-Hubble scales, with $k<<\hub$, the short-scale dynamics simply renormalize the background pressure, energy density, and anisotropic stress. We can determine the renormalization by evaluating $\sm{\tau_{\mu\nu}}$ as $k\to 0$.

The background energy density and pressure are given by
\begin{align}
\bar{\rho}_{\mathrm{eff}} = \lim_{k\to 0}-\sm{\tau^0_0}\hspace{2cm}\bar{P}_{\mathrm{eff}} = \lim_{k\to 0}\frac{1}{3}\sm{\tau^i_i}
\end{align}
The background anisotropic stress is zero. 

\subsection{Density and Velocity Perturbations}

To find the equations governing the dynamics of the effective fluid, we smooth the equations (\ref{eq-Cons1}-\ref{eq-Cons2}). Smoothing the continuity equation gives:
\begin{align}
&\dot{\rho_l} + \partial_i((\rho_l + P_l)v^i_l) - v_l^i\partial_iP_l + 3(\hub-\dot{\Phi}_l)(\rho_l + P_l) + 2\vel_l\cdot\dot{\vel_l}(\rho_l+P_l)\nonumber\\
&\hspace{1cm}= -\partial_i\sm{(\rho_s+P_s)v_s^i} + \sm{v_s^i\partial_iP_s} + 3\sm{\dot{\Phi}_s(\rho_s+P_s)} - \sm{(\rho+P)\partial_0(v^2)}^s\label{eq-SmCont}
\end{align}
From now on we will omit the ``$l$'' subscript from the variables to avoid cluttering the notation. Keeping only the terms linear in the perturbations $\delta_l$ and $\vel_l$, subtracting the homogeneous equation and dividing by $\bar{\rho}$ gives
\begin{align}
\dot{\delta} = (1+w)(3\dot{\Phi} - \grad\cdot\vel) - 3\hub\delta\left ( c_s^2 - w\right )
\end{align}
To this order we can take $w = \frac{\bar{P}}{\bar{\rho}}$. Similarly, smoothing Eq.(\ref{eq-Cons2}) gives:
\begin{align}
\grad\cdot\dot{\vel} + \frac{c_a^2}{(1+w)}\grad\cdot\vel + \grad^2\Psi &= -\frac{1}{\bar{\rho}(1+w)}\partial_i\partial_j\sm{\tau_{ij}}^s
\end{align}
where $w = \frac{\bar{P}}{\bar{\rho}}$ as before and $c_a^2 = \frac{\dot{\bar{P}}}{\dot{\bar{\rho}}}$ is the adiabatic sound speed. (The details of this calculation are somewhat involved and are therefore given in Appendix \ref{sec-App2}.) We can  introduce the velocity potential $v$ such that $v_i = ik_iv$:
\begin{align}
\dot{v} + \frac{c_a^2}{(1+w)}v - \Psi &= -\frac{1}{\bar{\rho}(1+w)}\left(c_s^2\bar{\rho}\delta - c_{vis}^2\frac{\theta}{\hub}\right)
\end{align}
Once again, our results differ from those given in \cite{Baumann}. We believe that the differences are due to extra simplifications that are made in \cite{Baumann}. For example, Baumann et al. assume that the Newtonian approximation holds at scales $k^{-1}<<\Lambda^{-1}$.

The Einstein equations for the long-wavelength metric perturbations are given in Appendix \ref{app-EE2}.

\subsection{Scalar Field Perturbations}

Finally, we consider the long-wavelength scalar field perturbations by smoothing the Klein-Gordon equations. In order to simplify our calculations we will ignore the metric perturbations, which is a commonly used approximation when studying field dynamics during preheating\cite{DEFROST,Frag,Felder,Podolsky}. Separating the fields $\phi,\chi$ into homogeneous parts $\bar{\phi},\bar{\chi}$ and perturbations $\delta{\phi},\delta{\chi}$, and assuming that $\bar{\chi} = 0$, we obtain the following equations for the perturbations:
\begin{align}
&\ddot{\delta\phi} - \grad^2(\delta\phi) + 2\hub\dot{\delta\phi} + a^2m^2\delta\phi + a^2g^2]\delta\chi^2(\bar{\phi}+\delta\phi) = 0\\
&\ddot{\delta\chi} - \grad^2(\delta\chi) + 2\hub\dot{\delta\chi} + a^2g^2\chi(\bar{\phi}+\delta\phi)^2 = 0
\end{align}
Smoothing these equations and keeping only the terms linear in the long-wavelength perturbations gives:
\begin{align}
&\ddot{\delta\phi_l} - \grad^2(\delta\phi_l) + 2\hub\dot{\delta\phi_l} + a^2m^2\delta\phi_l = -a^2g^2\left(\bar{\phi}\sm{\delta\chi_s^2} + \sm{\delta\chi_s^2\delta\phi_s}\right )\\
&\ddot{\delta\chi_l} - \grad^2(\delta\chi_l) + 2\hub\dot{\delta\chi_l} + a^2g^2\bar{\phi}^2\delta\chi_l = -a^2g^2\left(2\bar{\phi}\sm{\delta\chi_s\delta\phi_s} + \sm{\delta\chi_s\delta\phi_s^2}\right )
\end{align}
We have obtained evolution equations for $\delta\phi_l$ and $\delta\chi_l$ in terms of the 2 and 3-point correlation functions of the short-wavelength field perturbations. It would be interesting to find even approximate analytical forms for the correlation functions, so that the equations could be solved explicitly.

\section{Conclusion}\label{sec-Discussion}

We have studied field dynamics during preheating in an unusual way, by focusing our attention on long-wavelength perturbations and integrating out the short-wavelength modes. As might be expected, we see that the homogeneous part of the inflaton field plays an important part in the evolution of both the $\phi$ and $\chi$ fields, as do the correlation functions between the perturbations $\delta\phi$ and $\delta\chi$. We have also adapted and extended the formalism in \cite{Baumann} so that it can be applied to a universe whose matter content has pressure, and where the short-scale gravitational dynamics are not Newtonian. In contrast to the work in \cite{Baumann}, the only restriction we placed on our model was that the matter content should be modeled as a perfect fluid.

There are several obvious ways in which this work could be extended. Firstly, of course, it would be desirable to find analytic solutions of the equations of motion given in this paper, and to compare them with the results of lattice simulations of field dynamics during preheating. We have also given the linearized Einstein equations, which govern the evolution of large-scale scalar metric perturbations during preheating. Given that data collection is currently underway to find evidence distinguishing between various preheating models, it would be interesting to find ways of calculating observable non-Gaussianities in the power spectrum of the metric perturbations using this formalism. Also, we have only considered {\it scalar} metric perturbations in this work: the next step might be to consider the evolution of large-scale vector and tensor perturbations.

\begin{appendix}

\section{Smoothing Bilinear and Trilinear Quantities}\label{sec-App1}

Here we explicitly outline the calculations that lead to Eq.(\ref{eq-bi}-\ref{eq-tri}). Smoothing a bilinear quantity $fg$ gives the same result as in \cite{Baumann}, but we find a different expression for $\sm{fgh}$. Therefore we will follow the approach given in \cite{Baumann} to derive $\sm{fg}$, before outlining our calculation for $\sm{fgh}$ and emphasizing the differences from the result in \cite{Baumann}.

Recall that we assumed the following useful properties for the Gaussian window function $\win(|\textbf{x}-\textbf{x}'|)$:
\begin{align}
&\int_{\textbf{x}'} \win (\textbf{x}'-\textbf{x})^i(\textbf{x}'-\textbf{x})^j = \frac{1}{\Lambda^2}\delta^{ij}\label{eq-Win1}\\
&\partial_{j'}\win = -\partial_j\win = \Lambda^2(\textbf{x} - \textbf{x}')^j\win\label{eq-Win2}\\
&\partial_{i'}\partial_{j'}\win = \partial_i\partial_j\win = -\Lambda^2\delta_{ij}\win + \Lambda^4(\textbf{x} - \textbf{x}')^i(\textbf{x} - \textbf{x}')^j\win\label{eq-Win3}
\end{align}
To smooth $fg$, we take its convolution with the window function:
\begin{equation}
\sm{fg} = \int_{\textbf{x}'}\win f(\textbf{x}')g(\textbf{x}').
\end{equation}
Splitting the fields $f$, $g$ into long-wavelength modes $f_l, g_l$ and short-wavelength modes $f_s$, $g_s$, we get:
\begin{equation}
\sm{fg} = \sm{f_lg_l} + \sm{f_sg_s} + \sm{f_lg_s} + \sm{f_sg_l}.
\end{equation}
Since $f_l$ and $g_l$ are assumed to be small perturbations, and are long-scale, we can expand them in a Taylor series about $\textbf{x}$:
\begin{equation}
f_l(\textbf{x}') = f_l(\textbf{x})+\partial_if_l(\textbf{x})(\textbf{x}' - \textbf{x})^i + \frac{1}{2}\partial_i\partial_jf_l(\textbf{x})(\textbf{x}'-\textbf{x})^i(\textbf{x}'-\textbf{x})^j+\dots
\end{equation}
and similarly for $g_l$. This gives us
\begin{align}
\sm{f_lg_l} = f_lg_l + \frac{1}{\Lambda^2}\left (\grad f_l\cdot \grad g_l + \frac{1}{2}f_l\grad^2 g_l + \frac{1}{2}g_l\grad^2f_l \right ) + \dots\label{eq-LLSmooth}
\end{align}
where the dots indicate higher derivative terms, suppressed by powers of $k^2/\Lambda^2$.
To simplify the term $\sm{f_lg_s}$, we first rewrite it as
\begin{equation}
\sm{f_lg_s} = \sm{f_lg} - \sm{f_lg_l},
\end{equation}
where the second term is given by (\ref{eq-LLSmooth}). To simplify the first term, we again use the Taylor expansion of $f_l$, giving us:
\begin{align}
\sm{f_lg} &= f_lg_l - \partial_i f_l \sm{(\textbf{x}-\textbf{x}')^ig(\textbf{x}')} + \frac{1}{2}\partial_i\partial_j f_l \cdot \sm{(\textbf{x}-\textbf{x}')^i(\textbf{x}-\textbf{x}')^jg(\textbf{x}')} + \dots\nonumber\\
&= f_lg_l + \frac{1}{\Lambda^2} \left (\grad f_l\cdot\grad g_l + \frac{1}{2}g_l\grad^2 f_l \right )+\dots
\end{align}
Interchanging $f$ and $g$ in the above expression gives us $\sm{f_sg_l}$. Thus we find that
\begin{equation}
\sm{fg} = f_lg_l + \sm{f_sg_s} + \frac{1}{\Lambda^2}\grad f_l\cdot \grad g_l + \dots\label{eq-biApp}
\end{equation}

Smoothing a trilinear term $fgh$ proceeds in the same way. We begin by splitting the fields $f,g,h$ into long-wavelength modes $f_l,g_l,h_l$ and short-wavelength modes $f_s,g_s,h_s$, giving:
\begin{align}
\sm{fgh} &= \sm{f_lg_lh_l} + \sm{f_sg_sh_s} + \sm{f_lgh} +\sm{fg_lh} + \sm{fgh_l}\nonumber\\
&\hspace{4cm} - \sm{fg_lh_l} - \sm{f_lgh_l} - \sm{f_lg_lh}\label{eq-A3}
\end{align}
Expanding $f_l,g_l$ in Taylor series as before, we can smooth $f_lg_lh$:
\begin{align}
\sm{f_lg_lh} &= f_lg_lh_l - g_l\partial_i f_l \sm{(\textbf{x}-\textbf{x}')^ih(\textbf{x}')} - f_l\partial_i g_l\sm{(\textbf{x}-\textbf{x}')^ih(\textbf{x}')}\nonumber\\
&\hspace{1cm}\partial_i f_l\partial_j g_l \sm{(\textbf{x}-\textbf{x}')^i(\textbf{x}-\textbf{x}')^jh}\nonumber\\
&\hspace{1cm}\frac{1}{2}g_l\partial_i\partial_j f_l \sm{(\textbf{x}-\textbf{x}')^i(\textbf{x}-\textbf{x}')^jh} + \frac{1}{2}\partial_i\partial_j g_l\sm{(\textbf{x}-\textbf{x}')^i(\textbf{x}-\textbf{x}')^jh} + \dots\nonumber\\
&= f_lg_lh_l + g_l\frac{\grad f_l\cdot\grad h_l}{\Lambda^2} + f_l \frac{\grad g_l\cdot\grad h_l}{\Lambda^2} + h_l\frac{\grad f_l \cdot \grad g_l}{\Lambda^2}\nonumber\\
&\hspace{1cm}+ g_lh_l\frac{\grad^2 f_l}{2\Lambda^2} + f_lh_l\frac{\grad^2 g_l}{2\Lambda^2} + \dots\label{eq-A2}
\end{align}
We can also smooth $f_lgh$:
\begin{align}
\sm{f_lgh} &= f_l\sm{gh} - \partial_j f_l \sm{(\textbf{x}-\textbf{x}')^jgh}+\frac{1}{2}\partial_i\partial_j f_l\sm{(\textbf{x}-\textbf{x}')^i(\textbf{x}-\textbf{x}')^jgh}+\dots\nonumber\\
&= f_lg_lh_l + f_l\sm{g_sh_s} + f_l\frac{\grad g_l\cdot\grad h_l}{\Lambda^2}+g_l\frac{\grad f_l\cdot\grad h_l}{\Lambda^2}+h_l\frac{\grad f_l\cdot\grad g_l}{\Lambda^2}\nonumber\\
&\hspace{1cm}+g_lh_l\frac{\grad^2 f_l}{2\Lambda^2} + \frac{\grad f_l\cdot \grad\sm{g_sh_s}}{\Lambda^2} + \frac{\grad^2 f_l}{2\Lambda^2} \sm{g_sh_s}+ \dots,\label{eq-A1}
\end{align}
where we have used (\ref{eq-biApp}). Substituting (\ref{eq-A1}) and (\ref{eq-A2}) into (\ref{eq-A3}), we find
\begin{align}
\sm{fgh} &= f_lg_lh_l + \sm{f_sg_sh_s}+f_l\sm{g_sh_s}+g_l\sm{f_sh_s}+h_l\sm{f_sg_s}\nonumber\\
&+ \frac{1}{\Lambda^2}(f_l \grad g_l\cdot \grad h_l+g_l \grad f_l\cdot \grad h_l+h_l \grad f_l\cdot \grad g_l)\nonumber\\
&+ \frac{1}{\Lambda^2}(\grad f_l \cdot \grad\sm{g_sh_s}+\grad g_l \cdot \grad\sm{f_sh_s}+\grad h_l \cdot \grad\sm{f_sg_s})\nonumber\\
&+ \frac{1}{2\Lambda^2}(\grad^2 f_l \sm{g_sh_s}+\grad^2 g_l \sm{f_sh_s}+\grad^2 h_l \sm{f_sg_s})+ \dots\label{eq-triApp}
\end{align}
The expression easily generalizes to smoothed polynomial quantities $f_1f_2\cdots f_n$ of any order $n$.

Baumann et al. claim in \cite{Baumann} that smoothing the trilinear quantity $\rho v^iv^j$ gives:
\begin{align}
\sm{\rho v^iv^j} &= \rho_lv^i_lv^j_l + \sm{\rho v_s^iv_s^j} + \rho_l\frac{\grad v_l^i\cdot \grad v_l^j}{\Lambda^2} + \dots\nonumber\\
&= \rho_lv^i_lv^j_l+\sm{\rho_s v_s^iv_s^j} + \sm{\rho_l v_s^iv_s^j} + \rho_l\frac{\grad v_l^i\cdot \grad v_l^j}{\Lambda^2}+ \dots\nonumber\\
&= \rho_lv^i_lv^j_l +\sm{\rho_s v_s^iv_s^j}+ \rho_l\sm{v_s^iv_s^j} +\frac{\grad \rho_l\cdot \grad\sm{v_s^i v_s^j}}{\Lambda^2}+ \rho_l\frac{\grad v_l^i\cdot \grad v_l^j}{\Lambda^2}+ \dots
\end{align}
Comparing this to (\ref{eq-triApp}), we see that our general expression for a smoothed trilinear quantity $\sm{fgh}$ does not agree with Baumann et al.'s expression for $\sm{\rho v^i v^j}$ upon the substitution of $\rho, v^i,$ and $v^j$ for $f,g,$ and $h$, due to the presence of extra terms such as $v^i_l\sm{\rho_s v_s^j}$ in our formula. As $\sm{fgh}$ should be symmetrical in $f,g,h$, we believe that these terms should be present unless some additional assumptions or conditions are imposed. In our work, we will make no such additional assumptions.

\section{The Einstein Equations}\label{app-EE}

\subsection{Velocity Expansion of the Einstein Equations}\label{app-EE1}

As explained in Section \ref{sec-EE}, the energy-momentum pseudotensor $\tau_{\mu\nu}$ is given by
\begin{equation}
\tau_{\mu\nu} \equiv T_{\mu\nu} - \frac{G_{\mu\nu}^{NL}}{8\pi G}
\end{equation}
We expand these equations to order $v^2$. The expansion of $T_{\mu\nu}$ is given by Eq.(\ref{eq-EMT00}-\ref{eq-EMTij}). As explained in Section \ref{sec-Exp}, the metric perturbation $\Phi$ is of order $v^2$. However, each gradient $\grad\Phi$ lowers the order in $v$ by one, so terms of the form $\Phi\grad^2\Phi$ are of order $v^2$. We therefore work to first order in $\Phi$, with the exception of such gradient terms. We end up with the equations
\begin{align}
\grad^2\Phi - 3\hub(\dot{\Phi} + \hub\Psi) &=-4\pi G a^2(\tau^0_0 - \bar{T}^0_0)\label{eq-Eeq1}\\
\partial_i (\dot{\Phi} + \hub\Psi) &=4\pi G a^2\tau^i_0\label{eq-Eeq2}\\
\ddot{\Phi} + \hub(2\dot{\Phi}+\dot{\Psi}) + (\hub^2 + 2\dot{\hub})\Psi-\frac{2}{3}\grad^2(\Phi - \Psi) &=\frac{4\pi G a^2}{3}(\tau^i_i - \bar{T}^i_i)\label{eq-Eeq3}\\
\partial_i\partial_j\left [ \partial_i\partial_j(\Phi - \Psi) - \frac{1}{3}\delta_{ij}\grad^2(\Phi - \Psi)\right] &= 8\pi Ga^2\partial_i\partial_j (\tau_j^i - \frac{1}{3}\delta^i_j\tau^k_k)\label{eq-Eeq4}
\end{align}
where the non-linear parts of the Einstein tensor are
\begin{align}
-a^2(G^0_0)^{NL} &\sim -\grad\Phi\cdot\grad\Phi + 4\Phi\grad^2\Phi\\
-a^2(G^i_0)^{NL} &\sim 0\\
-a^2(G^i_j)^{NL} &\sim \delta^i_j\grad\Phi\cdot\grad\Phi - 2\grad_i\Phi\grad_j\Phi
\end{align}
to order $v^2$.
The homogeneous Einstein equations are
\begin{align}
\hub^2 &= -\frac{8\pi Ga^2}{3}\bar{T}^0_0 = -\frac{8\pi Ga^2}{3}\bar{\rho}\\
\hub^2+2\dot{\hub} &= -\frac{8\pi Ga^2}{3}\bar{T}^i_i =  -\frac{8\pi Ga^2}{3}\bar{P}
\end{align}

\subsection{The Einstein Equations in the Long-Wavelength Effective Theory}\label{app-EE2}

The long-wavelength metric perturbations are given by the Einstein equations, sourced by the effective energy-momentum pseudotensor. To linear order in the long-wavelength perturbations, we have
\begin{align}
&\grad^2\Phi_l - 3\hub(\dot{\Phi}_l + \hub\Psi_l) =-4\pi G a^2(\tau^0_0 - \bar{\tau}^0_0) = 4\pi G a^2\delta_l\bar{\rho}\\
&\partial_i (\dot{\Phi}_l + \hub\Psi_l) =4\pi G a^2\tau^i_0 = 4\pi G a^2(\bar{\rho}+\bar{P})v_l^i\\
&\ddot{\Phi}_l + \hub(2\dot{\Phi}_l+\dot{\Psi}_l) + (\hub^2 + 2\dot{\hub})\Psi_l-\frac{2}{3}\grad^2(\Phi_l - \Psi_l) =\frac{4\pi G a^2}{3}(\tau^i_i - \bar{\tau}^i_i)\nonumber\\
&\hspace{9cm}= \frac{4\pi G a^2}{3}c_s^2\bar{\rho}\delta_l\\
&\partial_i\partial_j\left [ \partial_i\partial_j(\Phi_l - \Psi_l) - \frac{1}{3}\delta_{ij}\grad^2(\Phi_l - \Psi_l)\right] = 8\pi Ga^2\partial_i\partial_j (\tau_j^i - \frac{1}{3}\delta^i_j\tau^k_k)\\
&\hspace{7cm} = 8\pi Ga^2(\bar{\rho} + \bar{P})c_{vis}^2\frac{\theta}{\hub}
\end{align}
where the components of $\tau^\mu_\nu$ are
\begin{align}
\tau^0_0 - \bar{\tau}^0_0 &= -\delta_l\rho_l - \sm{\rho v^2}^s - \frac{\sm{\Phi_{,k}^s\Phi_{,k}^s}-4\sm{\Phi^s\Phi_{,kk}^s}}{8\pi G a^2}\label{eq-Tau00}\\
\tau^i_0 &= (\rho_l + P_l)v^i_l + \sm{(\rho_s + P_s)v^i_s}\label{eq-Tau01}\\
\tau^i_j &= P_l\delta^i_j + \sm{(\rho+P)v^iv_j}^s - \frac{\sm{\Phi_{,k}^s\Phi_{,k}^s}\delta^i_j - 2\sm{\Phi_{,i}^s\Phi_{,j}^s}}{8\pi G a^2}\label{eq-Tauij}
\end{align}
Equations (\ref{eq-Tau00}-\ref{eq-Tauij}) differ slightly from those given in \cite{Baumann}, as we have different expressions for smoothed trilinear quantities. However, as the smoothed short-wavelength perturbations can be written as a derivative expansion of long-wavelength variables, we find that to leading order in the perturbation variables, the Einstein equations are the same.

\section{Smoothing the Momentum Conservation Equations}\label{sec-App2}

Here we show the explicit steps taken in smoothing the momentum conservation equation
\begin{equation}
(g^{\sigma\nu} + u^\sigma u^\nu)\grad_\mu\emt = (\rho + P)u^\mu \grad_\mu u^\sigma + \partial^\sigma P + u^\sigma u^\mu\partial_\mu P = 0
\end{equation}
which, when expanded to order $v^2$, becomes:
\begin{align}
(\rho + P)\left ( \dot{v_i} + v^j\partial_jv_i + \partial_i\Psi\right ) + \partial_i P + v_i\partial_0 P + v_iv^j\partial_j P = 0
\end{align}
We will omit all the higher derivative corrections suppressed by powers of $k^2/\Lambda^2$ that arise. Smoothing the term $\partial_iP$ is easy:
\begin{equation}
\sm{\partial_iP} = \partial_iP_l\label{eq-App1}
\end{equation}
Next we smooth the term $(\rho+P)\partial_i\Psi$:
\begin{align}
\sm{(\rho+P)\partial_i\Psi} &= (\rho_l + P_l)\partial_i\Psi_l + \sm{(\rho_s+P_s)\partial_i\Psi_s}\\
&= (\rho_l + P_l)\partial_i\Psi_l + \frac{1}{4\pi Ga^2}\sm{\grad^2\Psi_s\partial_i\Psi_s} + \sm{P_s\partial_i\Psi_s},
\end{align}
using the Poisson equation, $\grad^2\Psi_s = 4\pi G a^2\rho_s$. We then integrate by parts and use $\partial_{j'}\win =-\partial_j\win$ to get
\begin{align}
\sm{(\rho+P)\partial_i\Psi} &= (\rho_l + P_l)\partial_i\Psi_l + \sm{P_s\partial_i\Psi_s} + \frac{1}{8\pi Ga^2}\partial_j\sm{\Psi_{,k}\Psi_{,k}\delta^i_j - 2\Psi_{,i}\Psi_{,j}}\label{eq-App2}
\end{align}
Finally we smooth the remaining terms, integrating by parts and using the properties (\ref{eq-Win1}-\ref{eq-Win3}) of the window function when necessary:
\begin{align}
&\sm{(\rho + P)\left ( \dot{v_i} + v^j\partial_jv_i\right) + v_i\dot{P} + v_iv^j\partial_j P}\nonumber\\
&\hspace{0.5cm}=\partial_0\sm{(\rho+P)v^i} - \int\win v^i(\dot{\rho}+\dot{P}) + \int\win \left ( (\rho+P)v^j\partial_jv_i + v^i\dot{P} + v^iv_j\partial_jP\right )\nonumber\\
&\hspace{0.5cm}=\partial_0\sm{(\rho+P)v^i} -\int\win v^i\left (-\partial_{j'}((\rho+P)v^j) - 3(\hub-\dot{\Phi})(\rho+P) - (\rho+P)\dot(v^2) \right )\nonumber\\
&\hspace{1.5cm} + \int\win (\rho+P)v^j\partial_jv_i\nonumber\\
&\hspace{0.5cm}=\partial_0\sm{(\rho+P)v^i} + \int\win\left ( v^i\partial_{j'}((\rho+P)v^j)+(\rho+P)v^j\partial_jv_i\right )\nonumber\\
&\hspace{1.5cm} +\int\win v^i\left ( 3(\hub-\dot{\Phi})(\rho+P) +(\rho+P)\dot{(v^2)}\right )\nonumber\\
&\hspace{0.5cm}=\partial_0\sm{(\rho+P)v^i} + \partial_j\int\win(\rho + P)v^iv^j\nonumber\\
&\hspace{1.5cm} + \int\win v^i\left ( 3(\hub-\dot{\Phi})(\rho+P) +(\rho+P)\dot{(v^2)}\right )
\end{align}
Using the smoothed continuity equation (\ref{eq-SmCont}), we can substitute for $\dot{\rho}+\dot{P}$, which gives:
\begin{align}
&\sm{(\rho + P)\left ( \dot{v_i} + v^j\partial_jv_i\right) + v_i\dot{P} + v_iv^j\partial_j P}\nonumber\\
&\hspace{0.5cm}=(\rho_l+P_l)\dot{v_l}^i + \partial_0\sm{(\rho_s+P_s)v_s^i}+ \partial_j\int\win(\rho + P)v^iv^j\nonumber\\
&\hspace{1.5cm}+ v_l^i\left ( -\partial_j((\rho_l+P_l)v_l^j)-\partial_j\sm{(\rho_s+P_s)v_s^i}+\dot{P_l}+v_l^j\partial_jP_l + \sm{v_s^j\partial_jP_s}\right )\label{eq-App3}
\end{align}
Adding Eq.(\ref{eq-App1}-\ref{eq-App3}) gives:
\begin{align}
&(\rho_l+P_l)\dot{v_l}^i + (\rho_l+P_l)v_l^j\partial_jv_i^l + v_i^l\dot{P}_l + v_l^iv_l^j\partial_jP_l +(\rho_l + P_l)\partial_i\Psi_l+\partial_i P_l\nonumber\\
&\hspace{0.5cm}= \partial_j\left[\frac{\Psi_{,k}\Psi_{,k}\delta^i_j - 2\Psi_{,i}\Psi_{,j}}{8\pi Ga^2}\right]_{\Lambda} + v_l^i\partial_j\sm{(\rho_s+P_s)v_s^i} - v_l^i\sm{v_s^j\partial_j P_s}\nonumber\\
&\hspace{1.5cm} - \partial_0\sm{(\rho_s+P_s)v_s^i} - \partial_j\sm{(\rho+P)v^iv^j}^s - \sm{P_s\partial_i\Psi_s}
\end{align}
Keeping only the leading terms linear in the matter and velocity perturbations, we get:
\begin{align}
(1+w)\dot{v_l}^i + v^i_l\frac{\dot{P}_l}{\rho_l} + (1+w)\partial_i\Psi_l + \frac{\partial_i P_l}{\rho} = \partial_j\sm{\tau_{ij}}^s
\end{align}

\end{appendix}

\end{document}